\documentclass[preprint,aps]{revtex4}[10pt]
\usepackage{graphicx}
\usepackage{amsmath}
\usepackage{bm}
\draft
\begin{document}
\title{Modulational interactions in quantum plasmas}
\author{F. Sayed$^{1}$, S. V. Vladimirov$^{1,2,3}$, Yu. Tyshetskiy$^{1}$, and O. Ishihara$^{2}$}
\address{$^{1}$School of Physics, University of Sydney, New South Wales 2006, Australia\\
$^{2}$Faculty of Engineering, Yokohama National University, Yokohama 240-8501, Japan\\
$^{3}$Metamaterials Laboratory, National Research University of Information Technology,
Mechanics, and Optics, St Petersburg 199034, Russia
}
\begin{abstract}
A formalism for treating modulational interactions of electrostatic fields
in collisionless quantum plasmas is developed, based on the
kinetic Wigner-Poisson model of quantum plasma. This
formalism can be used in a range of problems of nonlinear interaction 
between electrostatic fields in a quantum plasma, such as development
of turbulence, self-organization, as well as transition from the
weak turbulent state to strong turbulence. In particular, using
this formalism, we obtain the kinetic quantum Zakharov
equations, that describe nonlinear coupling of high frequency Langmuir 
waves to low frequency plasma density variations, for cases
of non-degenerate and degenerate plasma electrons.
\end{abstract}
\maketitle
\begin{sloppypar}
\section{Introduction}
Modern plasma physics deals mainly with nonlinear phenomena, 
which are often dominant in basic plasma research as well as in many 
experimental and industrial applications. The applications of 
nonlinear plasma physics are wide-ranging, and plasma nonlinear effects 
are often used to illustrate general nonlinear phenomena in arbitrary 
media. One of the specific nonlinear phenomena in plasma physics is
modulational interaction~\cite{r1,r2}. Such interactions describe various 
modulational effects in nonlinear media, such as amplitude modulation, 
frequency modulation, phase modulation, self modulation, etc.~\cite{r3}.
Modulational phenomena play a key role in  the development of many nonlinear 
plasma processes. The process of 
modulational interactions is especially significant for high power energy 
input into a plasma as in, e.g., the development of turbulence, the process of
self-organization, and the transition from the weak
turbulent state to strong turbulence. These phenomena result in
the formation of strongly correlated structures (solitons,
cavitons, etc.), the generation of strong magnetic fields, heating,
and effective particle acceleration. Because of the important role of 
modulational interactions in plasma physics, many works are devoted to them
~\cite{r1,r2,r3,r4,r5,r6,r7,r8,r9,r10}.

Recently, there has been increasing interest in quantum plasmas
due to their relevance to modern laser-matter interaction experiments (e.g., the 
compressed hydrogen in the fast ignition scenario of inertial fusion is in a quantum 
plasma state), as well as their ubiquity
in different astrophysical and cosmological
systems~\cite{r11,r12,r13} (e.g., interstellar or molecular clouds,
planetary rings, comets, interiors of white dwarf stars, etc.), in
nanostructures~\cite{r14}, and in microelectronic devices~\cite{r15}. Many
authors include quantum corrections to quantum plasma
echoes~\cite{r16}, self-consistent dynamics of Fermi
gases~\cite{r17}, quantum beam instabilities~\cite{r18},
dispersion of ion acoustic waves~\cite{r19}, classical and
quantum kinetics of the Zakharov system~\cite{r20}, quantum
corrections to the Zakharov equations~\cite{r21}, expansion of
quantum electron gases into vacuum~\cite{r22}, quantum ion acoustic
waves~\cite{r23}, quantum Landau damping~\cite{r24},
magnetohydrodynamics of quantum plasmas~\cite{r25}, etc. Quantum
plasmas have extremely high plasma number densities and/or low
temperatures. At low temperatures, the thermal de-Broglie 
wavelength becomes comparable to the inter-electron
distance and the electron temperature becomes comparable to the
electron Fermi temperature $T_{F}$, defined as 
\begin{eqnarray}
&&\hspace{-2.8cm}k_{B}T_F\equiv E_F=\frac{\hbar^2}{2m}(3\pi^2)^{2/3}n^{2/3}, \label{q1}
\end{eqnarray}
and the electron energy distribution becomes step-like (degenerate limit). In this case, 
quantum mechanical effects play a significant role in the behavior
of charged particles~\cite{r26,r27,r28,r29,r30}. As electrons are
lighter than ions, the quantum behavior of electrons is reached
faster than ions. At room temperature and standard metallic
densities, the electron gas in an ordinary metal is a good
example of a quantum plasma system. The concept of quantum plasma
is also applicable in semiconductor physics. The electron density
in semiconductors is much lower than in metals, but the great
degree of miniaturization of today's electronic components is
such that the de Broglie wavelength of the charge carriers can be
comparable to the spatial variation of the doping profiles. In
the behavior of such electronic components, typical quantum
mechanical effects (e.g., quantum tunneling effects) are expected
to play a central role. Another possible application of quantum
plasmas arises from astrophysics. In astrophysical and
cosmological compact objects, the density of charged particles is
extremely high (some ten orders of magnitude larger than that of
ordinary solids). The properties of plasma existing in such
ultra-dense states possess strong quantum effects and exhibit
fluid and crystal properties in a quantum sea of
electrons~\cite{r12}.

The theory of modulational interactions in classical plasmas,
based on the formalism of nonlinear kinetic classical plasma
response, is well developed~\cite{r4}. It provides a systematic
description of different modulational interaction processes,
including the nonlinear coupling of high frequency Langmuir waves
with low frequency plasma density variations that is described by
the well-known Zakharov equations for classical
plasmas~\cite{r31}. In this paper, we generalize the formalism of
modulational interactions to nonrelativistic quantum plasmas,
based on the Wigner kinetic description of collisionless quantum
plasmas. In particular, we derive kinetically the effective cubic
response of a quantum plasma (which in general is a
complex-valued function), which can be used for various
modulational processes. As an illustration of its use, we derive
the quantum-corrected Zakharov equations for collisionless
quantum plasmas by neglecting the imaginary part of the effective
cubic response.

Our paper is organized as follows. In Sec. II we write the
quantum kinetic model for the plasma particles, derive the
higher order distribution functions, and thus obtain the formalism for
modulational interactions in quantum plasmas. 
In Sec. III we derive nonlinear responses for both non-degenerate and 
degenerate plasma electrons. In Sec. IV we
apply this formalism and derive the kinetic quantum Zakharov
equations for both non-degenerate and degenerate plasma electrons.
Finally, in Sec. V we summarize our results and discuss their
physical significance.
\section{Formalism for modulational interactions}
We start with the Wigner kinetic equation~\cite{r32,r33} for the 
quantum electron distribution function (Wigner function) $f_e$ that reads~\cite{r32}
\begin{eqnarray}
\frac{\partial f_e}{\partial t}+{\bf
v}\cdot\nabla f_e&=&-\frac{iem_{e}^3}{(2\pi)^3\hbar^4}\int \int d{\bm{\lambda}} d{\bf v}'\exp\left[i\frac{m_{e}}{\hbar}
({\bf v}-{\bf v}')\cdot{\bm{\lambda}}\right]\nonumber\\
&\times&\left[\phi\left(\mathbf{x}+\frac{\bm{\lambda}}{2},t\right)-\phi\left(\mathbf{x}-
\frac{{\bm{\lambda}}}{2},t\right)\right]
f_e(\mathbf{x},{\bf v}',t), \label{q2}
\end{eqnarray}
where $\phi$ is the electric potential obeying Poisson's equation
\begin{eqnarray}
&&\hspace{-2.8cm}\nabla\cdot{\bf E}=4\pi\sum_{e,i} q_{e,i}\int
\frac{d{\bf p}}{(2\pi\hbar)^3}f_{e,i}, \label{q3}
\end{eqnarray}
where ${\bf E}=-\nabla\phi$, $q_{e,i}=(\mp)e$ with $e$ the
magnitude of the electron charge, $m_e$ is the electron mass,
$\hbar$ is the reduced Planck constant and $f_i$ is the ion distribution 
function. We are interested in the plasma response to the 
self-consistent electrostatic field $E$. To describe this response, we need to 
solve the kinetic equation to find $f_e$. The solution can be 
sought in the form of the expansion
\begin{eqnarray}
&&\hspace{-2.8cm}f_e=f_e^{(0)}+f_e^{(1)}+f_e^{(2)}+\dots,
\label{q4}
\end{eqnarray}
where $f_e^{(0)}$ is the unperturbed distribution function which is assumed to be uniform (i.e., independent of {\bf r}), and $f_e^{(1)}$, $f_e^{(2)}$ etc. are the small perturbations (linear,
quadratic, etc.) of the electron distribution
function in powers of the electric field strength, i.e., $ f_e^{(n)}\propto E^n$. 
We consider weak fields, linearize Eq.~(\ref{q2}) using the weak field approximation 
(i.e. $|f_e^{(0)}|\gg |f_e^{(1)}|\gg
|f_e^{(2)}|\gg\dots$),
and Fourier transform Eqs.~(\ref{q2})~and~(\ref{q3}). The resulting linearized equation 
for $\tilde{f}_e({\bf k},\omega)$ and $\phi({\bf k},\omega)$ - the Fourier transforms of the corresponding functions $f_e({\bf r},t)$ and $\phi({\bf r},t)$ - follows from Eq.~(\ref{q2}) as
\begin{eqnarray}
&&\hspace{-1.2cm}(\omega-{\bf k}\cdot{\bf
v})\tilde{f}_e^{(1)}=\frac{em_{e}^3}{(2\pi)^3\hbar^4}\int\int d{\bm{\lambda}} d{\bf v}'
\exp\left[i\frac{m_{e}}{\hbar}({\bf v}-{\bf v}')\cdot\bm {\lambda}\right]\nonumber\\
&&\hspace{2.8cm}\times\left[e^{i{\bf k}\cdot \frac{{\bm
{\lambda}}}{2}}-e^{-i{\bf k}\cdot \frac{{\bm
{\lambda}}}{2}}\right]\phi{({\bf
k},\omega)}f_e^{(0)}({\bf v}'),\label{q5}
\end{eqnarray}
which can be rewritten as
\begin{eqnarray}
&&\hspace{-1.2cm}(\omega-{\bf k}\cdot{\bf
v})\tilde{f}_e^{(1)}=\frac{em_{e}^3}{(2\pi)^3\hbar^4}\int\int d{\bm{\lambda}} d{\bf v}' \bigg\{\exp\left[i\frac{m_{e}}{\hbar}({\bf v}-{\bf
v}')\cdot{\bm {\lambda}}
+i{\bf k}\cdot\frac{{\bm {\lambda}}}{2}\right]\nonumber\\
&&\hspace{2.0cm}-\exp\left[i\frac{m_{e}}{\hbar}({\bf v}-{\bf
v}')\cdot{\bm {\lambda}}-i{\bf k}\cdot\frac{{\bm {\lambda}}}{2}\right]\bigg\}\phi{({\bf
k},\omega)}f_e^{(0)}({\bf v}').\label{q6}
\end{eqnarray}
Now integrating over $\bm {\lambda}$ -space gives
\begin{eqnarray}
&&\hspace{-1.4cm}(\omega-{\bf k}\cdot{\bf
v})\tilde{f}_e^{(1)}=\frac{em_{e}^3}{\hbar^4}\int
d{\bf v}'\bigg\{\delta\left[\frac{m_{e}}{\hbar}({\bf v}-{\bf v}')+\frac{{\bf k}}{2}\right]\nonumber\\
&&\hspace{1.8cm}-\delta\left[\frac{m_{e}}{\hbar}({\bf v}-{\bf
v}')-\frac{{\bf k}}{2}\right]\bigg\} \phi{({\bf
k},\omega)}f_e^{(0)}({\bf v}'),\label{q7}
\end{eqnarray}
where $\delta$ is the Dirac delta function. Now on integrating
over ${\bf v}'$ - space, we have
\begin{eqnarray}
&&\hspace{-2.8cm}\tilde{f}_e^{(1)}=\left(\frac{e}{\hbar}\right)\frac{1}{\Omega}
\Big[f_e^{(0)}({\bf v}+{\bf{\Delta}})-f_e^{(0)}({\bf
v}-{\bf{\Delta}})\Big]\phi{({\bf k},\omega)},\label{q8}
\end{eqnarray}
where ~${\Omega}= \omega-{\bf k}\cdot{\bf v}$, ${\bf{\Delta}}=\frac{\hbar{\bf{k}}}{2m_{e}}$.
\\
For the higher-order perturbations of the electron distribution functions,
 $f_e^{(n)}$ with $n\geq 2$, we use the convolution theorem and obtain the following results:
\begin{eqnarray}
&&\hspace{-0.5cm}\tilde{f}_e^{(2)}=\left(\frac{e}{\hbar}\right)^2\frac{1}{\Omega}\int
d_{12}\bigg(\Big\{\frac{1}{(\Omega_{2}-{\bf{\Delta}}_2\cdot{\bf k}_1)}
[f_e^{(0)}\left({\bf
v}+{\bf{\Delta}}_{1+2}\right)-f_e^{(0)}\left({\bf
v}+{\bf{\Delta}}_{1-2}\right)]\Big\}\nonumber\\
&&\hspace{1.5cm}~~~~~-\Big\{\frac{1}{(\Omega_{2}+{\bf{\Delta}}_2\cdot{\bf k}_1)}
[f_e^{(0)}\left({\bf
v}-{\bf{\Delta}}_{1+2}\right)-f_e^{(0)}\left({\bf
v}-{\bf{\Delta}}_{1-2}\right)]\Big\}\bigg)\nonumber\\
&&\hspace{9.1cm}\phi({\bf
k}_1,\omega_{1})\phi({\bf k}_2,\omega_{2}),\label{q9}
\end{eqnarray}
\begin{eqnarray}
&&\hspace{0.4cm}\tilde{f}_e^{(3)}=\left(\frac{e}{\hbar}\right)^3\frac{1}{\Omega}\int
d_{123}\Big(\frac{1}{(\Omega-\Omega_{1}-{\bf{\Delta}}\cdot{\textbf{k}}_1+{\Delta}_{1}^2)}
~~\Big\{\frac{1}{(\Omega_{3}-{\bf{\Delta}}_3\cdot{\textbf{k}}_1-{\bf{\Delta}}_3\cdot{\textbf{k}}_2)}
\nonumber\\
&&\hspace{0.5cm}~~~~~~~~[f_e^{(0)}\left({\bf
v}+{\bf{\Delta}}_{1+2+3}\right)-f_e^{(0)}\left({\bf
v}+{\bf{\Delta}}_{1+2-3}\right)]\Big\}
-\Big\{\frac{1}{(\Omega_{3}-{\bf{\Delta}}_3\cdot{\textbf{k}}_1+{\bf{\Delta}}_3\cdot{\textbf{k}}_2)}
\nonumber\\
&&\hspace{0.5cm}~~~~~~~~[f_e^{(0)}\left({\bf
v}+{\bf{\Delta}}_{1-2+3}\right)-f_e^{(0)}\left({\bf
v}+{\bf{\Delta}}_{1-2-3}\right)]\Big\}\Big)
-\Big(\frac{1}{(\Omega-\Omega_{1}+{\bf{\Delta}}\cdot{\textbf{k}}_1-{\Delta}_{1}^2)}
\nonumber\\
&&\hspace{0.5cm}~~~~~~~~\Big\{\frac{1}{(\Omega_{3}+{\bf{\Delta}}_3\cdot{\textbf{k}}_1-{\bf{\Delta}}_3\cdot{\textbf{k}}_2)}
[f_e^{(0)}\left({\bf
v}-{\bf{\Delta}}_{1+2+3}\right)-f_e^{(0)}\left({\bf
v}-{\bf{\Delta}}_{1+2-3}\right)]\Big\}\nonumber\\
&&\hspace{0.5cm}~~~~~-\Big\{\frac{1}{(\Omega_{3}+{\bf{\Delta}}_3\cdot{\textbf{k}}_1+{\bf{\Delta}}_3\cdot{\textbf{k}}_2)}
[f_e^{(0)}\left({\bf
v}-{\bf{\Delta}}_{1-2+3}\right)-f_e^{(0)}\left({\bf
v}-{\bf{\Delta}}_{1-2-3}\right)]\Big\}\Big)\nonumber\\
&&\hspace{8.2cm}\phi({\bf k}_1,\omega_{1})\phi({\bf
k}_2,\omega_{2})\phi({\bf k}_3,\omega_{3}),\label{q10}
\end{eqnarray}
where 
\begin{eqnarray} &&\hspace{-1.8cm}d_{12}=d\omega_1 ~d{\bf k}_1 ~d\omega_2 ~d{\bf k}_2
~\delta({\bf k}-{\bf k_{1}}-{\bf
k_{2}})\delta(\omega-\omega_{1}-\omega_{2}),\label{q11}
\end{eqnarray}
\begin{eqnarray} &&\hspace{-1.8cm}d_{123}=d\omega_1 ~d{\bf k}_1 ~d\omega_2
~d{\bf k}_2 ~d\omega_3 ~d{\bf k}_3 ~\delta({\bf k}-{\bf k_{1}}-{\bf k_{2}}-{\bf
k_{3}})\delta(\omega-\omega_{1}-\omega_{2}-\omega_{3}),\label{q12}
\end{eqnarray}
${\bf{\Delta}}_{1+2}={\bf{\Delta}}_1+{\bf{\Delta}}_2$, ${\bf{\Delta}}_{1-2}={\bf{\Delta}}_1-{\bf{\Delta}}_2$, ${\bf{\Delta}}_{1+2+3}={\bf{\Delta}}_1+{\bf{\Delta}}_2+{\bf{\Delta}}_3$,
${\bf{\Delta}}_{1+2-3}={\bf{\Delta}}_1+{\bf{\Delta}}_2-{\bf{\Delta}}_3$,
${\bf{\Delta}}_{1-2+3}={\bf{\Delta}}_1-{\bf{\Delta}}_2+{\bf{\Delta}}_3$,
${\bf{\Delta}}_{1-2-3}={\bf{\Delta}}_1-{\bf{\Delta}}_2-{\bf{\Delta}}_3$,

and

~${\Omega_j}= \omega_j-{\bf k}_j\cdot{\bf v}$, ~${\bf{\Delta}}_j=\frac{\hbar{\bf{k}}_j}{2m_{e}}$,
for $j=1, 2,$ and $3$.
\\
According to the definition of the nonlinear responses, the
quadratic and cubic responses $S$ and $\Sigma$
are given by
\begin{eqnarray}
&&\hspace{-1.8cm}\frac{4\pi e}{i|{\bf k}|}\int\frac{2d{\bf
p}}{(2\pi\hbar)^3} \tilde{f}^{(2)}_e({{\bf k},\omega})=\int
d_{12}S_{1,2}E_{1}E_{2},\label{q13}
\end{eqnarray}
and
\begin{eqnarray}
&&\hspace{-1.8cm}\frac{4\pi e}{i|{\bf k}|}\int\frac{2d{\bf
p}}{(2\pi\hbar)^3} \tilde{f}^{(3)}_e({{\bf k},\omega})=\int
d_{123}\Sigma_{1,2,3}E_{1}E_{2}E_{3}.\label{q14}
\end{eqnarray}
We symmetrize the nonlinear responses obtained from Eqs.~(\ref{q13}) and (\ref{q14}) and write
\begin{eqnarray}
&&\hspace{-1.0cm}S_{1,2}=\int \frac{2 d{\bf p}}{(2\pi
\hbar)^3}\frac{2\pi i e^3}{\hbar^2}
\times\bigg[\frac{1}{\Omega+{\bf{\Delta}}\cdot(\textbf{k}_1+\textbf{k}_2)}
~\bigg\{\frac{1}{(\Omega_1+{\Delta}_1^2)}+\frac{1}
{(\Omega_2+{\Delta}_2^2)}\bigg\}\nonumber\\
&&\hspace{0.4cm}+\frac{1}{\Omega-{\bf{\Delta}}\cdot(\textbf{k}_1+\textbf{k}_2)}
\bigg\{\frac{1}{(\Omega_1-{\Delta}_1^2)}+\frac{1}
{(\Omega_2-{\Delta}_2^2)}\bigg\}-\frac{1}
{\Omega+{\bf{\Delta}}\cdot(\textbf{k}_1-\textbf{k}_2)}\nonumber\\
&&\hspace{0.6cm}\bigg\{\frac{1}{(\Omega_1+{\Delta}_1^2)}+\frac{1}
{(\Omega_2-{\Delta}_2^2)}\bigg\}-\frac{1}
{\Omega-{\bf{\Delta}}\cdot(\textbf{k}_1-\textbf{k}_2)}
~\bigg\{\frac{1}{(\Omega_1-{\Delta}_1^2)}\nonumber\\
&&\hspace{5.8cm}+\frac{1}
{(\Omega_2+{\Delta}_2^2)}\bigg\}\bigg]
\times \frac{f_e^{(0)}({\bf v})}{|{\bf
k}_1||{\bf k}_2||{\bf k}_1+{\bf
k}_2|}, \label{q15}
\end{eqnarray}
and
\begin{eqnarray}
&&\hspace{-0.8cm}\Sigma_{1,2,3}=\int \frac{2 d{\bf p}}{(2\pi
\hbar)^3}\frac{2\pi e^4}{\hbar^3}
\times
\bigg[\frac{1}{\Omega+{\bf{\Delta}}\cdot(\textbf{k}_1+\textbf{k}_2+\textbf{k}_3)}
\bigg\{\frac{1}
{\Omega-\Omega_{1}+{\bf{\Delta}}\cdot({\textbf{k}}_2+{\textbf{k}}_3)-{\bf{\Delta}}_1\cdot({\textbf{k}}_2+
{\textbf{k}}_3)}\nonumber\\
&&\hspace{0.3cm}\bigg\{\frac{1}{(\Omega_3+{\Delta}_3^2)}+\frac{1}{(\Omega_2+{\Delta}_2^2)}\bigg\}\bigg\}
+\bigg\{\frac{1}{\Omega+{\bf{\Delta}}\cdot(\textbf{k}_1-\textbf{k}_2-\textbf{k}_3)}\bigg\{\frac{1}{(\Omega_3-{\Delta}_3^2)}+\frac{1}{(\Omega_2-{\Delta}_2^2)}\bigg\}\nonumber\\
&&\hspace{0.5cm}\frac{1}
{\Omega-\Omega_{1}-{\bf{\Delta}}\cdot({\textbf{k}}_2+{\textbf{k}}_3)+{\bf{\Delta}}_1\cdot({\textbf{k}}_2+{\textbf{k}}_3)}\bigg\}
-\bigg\{\frac{1}{\Omega+{\bf{\Delta}}\cdot(\textbf{k}_1+\textbf{k}_2-\textbf{k}_3)}\bigg\{\frac{1}{(\Omega_3-{\Delta}_3^2)}\nonumber\\
&&\hspace{0.3cm}+\frac{1}{(\Omega_2+{\Delta}_2^2)}\bigg\}\frac{1}
{\Omega-\Omega_{1}+{\bf{\Delta}}\cdot({\textbf{k}}_2-{\textbf{k}}_3)-{\bf{\Delta}}_1\cdot({\textbf{k}}_2
-{\textbf{k}}_3)}\bigg\}
-\bigg\{\frac{1}{\Omega+{\bf{\Delta}}\cdot(\textbf{k}_1-\textbf{k}_2+\textbf{k}_3)}
\nonumber\\
&&\hspace{0.3cm}\bigg\{\frac{1}{(\Omega_3+{\Delta}_3^2)}+\frac{1}{(\Omega_2-{\Delta}_2^2)}\bigg\}\frac{1}
{\Omega-\Omega_{1}-{\bf{\Delta}}\cdot({\textbf{k}}_2-{\textbf{k}}_3)+{\bf{\Delta}}_1\cdot({\textbf{k}}_2-{\textbf{k}}_3)}\bigg\}-\bigg\{\frac{1}{(\Omega_3+{\Delta}_3^2)}
\nonumber\\
&&\hspace{0.3cm}+\frac{1}{(\Omega_2+{\Delta}_2^2)}\bigg\}
\bigg\{\frac{1}{\Omega-{\bf{\Delta}}\cdot(\textbf{k}_1-\textbf{k}_2-\textbf{k}_3)}\bigg\{\frac{1}
{\Omega-\Omega_{1}+{\bf{\Delta}}\cdot({\textbf{k}}_2+{\textbf{k}}_3)-{\bf{\Delta}}_1\cdot({\textbf{k}}_2+{\textbf{k}}_3)}
\bigg\}\bigg\}
\nonumber\\
&&\hspace{0.5cm}-\frac{1}{\Omega-{\bf{\Delta}}\cdot(\textbf{k}_1+\textbf{k}_2+\textbf{k}_3)}
~\bigg\{\frac{1}
{\Omega-\Omega_{1}-{\bf{\Delta}}\cdot({\textbf{k}}_2+{\textbf{k}}_3)+{\bf{\Delta}}_1\cdot({\textbf{k}}_2+{\textbf{k}}_3)}
~\bigg\{\frac{1}{(\Omega_3-{\Delta}_3^2)}\nonumber\\
&&\hspace{0.3cm}+\frac{1}{(\Omega_2-{\Delta}_2^2)}\bigg\}\bigg\}
+\frac{1}{\Omega-{\bf{\Delta}}\cdot(\textbf{k}_1+\textbf{k}_2-\textbf{k}_3)}
\bigg\{\frac{1}
{\Omega-\Omega_{1}-{\bf{\Delta}}\cdot({\textbf{k}}_2-{\textbf{k}}_3)+{\bf{\Delta}}_1\cdot({\textbf{k}}_2-{\textbf{k}}_3)}
\nonumber\\
&&\hspace{0.3cm}\bigg\{\frac{1}{(\Omega_3+{\Delta}_3^2)}+\frac{1}{(\Omega_2-{\Delta}_2^2)}\bigg\}\bigg\}
+\bigg\{\frac{1}{\Omega-{\bf{\Delta}}\cdot(\textbf{k}_1-\textbf{k}_2+\textbf{k}_3)}\bigg\{\frac{1}{(\Omega_3-{\Delta}_3^2)}+\frac{1}{(\Omega_2+{\Delta}_2^2)}\bigg\}
\nonumber\\
&&\hspace{2.0cm}\frac{1}
{\Omega-\Omega_{1}+{\bf{\Delta}}\cdot({\textbf{k}}_2-{\textbf{k}}_3)-{\bf{\Delta}}_1\cdot({\textbf{k}}_2-{\textbf{k}}_3)}\bigg\}\bigg]
\times \frac{f_e^{(0)}({\bf v})}{|{\bf
k}_1||{\bf k}_2||{\bf k}_3||{\bf k}_1+{\bf
k}_2+{\bf k}_3|}.\label{q16}
\end{eqnarray}
To obtain the evolution of high frequency field equations, we use the Poisson equation
(for longitudinal waves) in which the terms up to the third order in
electric field $E$ are taken into account. In Fourier components we have 
\begin{eqnarray}
&&\hspace{-0.2cm}\varepsilon E=\int
d_{12}S_{1,2}E_{1}E_{2}+ \int
d_{123}\Sigma_{1,2,3}E_{1}E_{2}E_{3},\label{q17}
\end{eqnarray}
where $\varepsilon=\varepsilon (\omega, {\bf k})$ is the linear dielectric permittivity of the plasma,
$S_{1,2}$ and $\Sigma_{1,2,3}$ are
the nonlinear plasma responses of the second and third order in field
$E$. The high frequency Langmuir wave is approximated under the
assumption that the phase speed is much greater than the electron
thermal speed (i.e. $\omega>>$max$(kv_{Te}, kv_{Fe})$). The approximation
of the Langmuir wave frequency is given by 
\begin{eqnarray}
&&\hspace{-0.2cm}\omega({\bf k})=\omega_{pe}+
C\frac{k^2v^2_{\sigma}}{\omega_{pe}},\label{q18}
\end{eqnarray}
where $C=3/2,\ \sigma=T_e$ for non-degenerate plasma electrons,
and $C=3/10,\ \sigma=T_F$ for fully degenerate plasma electrons. For $\omega>>$max$(kv_{Te}, kv_{Fe})$ the Langmuir wave frequency weakly depends on ${\bf k}$. Therefore we can include the waves with all possible wave vectors in field $E$ defined by Eq.~(\ref{q17}). Langmuir waves have frequency 
close to $\pm 1$ (for simplicity, we count here in units of the electron plasma frequency, 
$\omega_{pe}$), while the virtual wave fields~\cite{r34} have frequencies close to $0$, $\pm 2$, $\pm 3$ etc. The field of the virtual wave is real, and appears from the nonlinear response of the plasma to the real wave field. It is called the ``virtual wave field" because it has a certain frequency and wave vector, dictated by the laws of energy $(\omega)$ and momentum $(\bf{k})$ conservation in multi-wave interactions in the nonlinear medium. Yet its frequency and wave vector are not related by any dispersion relation of any natural oscillation of the medium (plasma), hence the name ``virtual". In other words, the virtual wave fields are the fields of forced oscillations of the plasma, driven by the interacting natural  electrostatic plasma modes (in our case by Langmuir modes), but are not the natural oscillations of the plasma themselves. 
The virtual wave field on the ``zero" frequency (which implies that their frequency is small compared to the Langmuir wave frequency) can be written in the form
\begin{eqnarray}
&&\hspace{-0.2cm}\varepsilon E^{0}=\int
d_{12}S_{1,2}E^{0}_{1}E^{0}_{2}+ \int
d_{123}\Sigma_{1,2,3}E^{0}_{1}E^{0}_{2}E^{0}_{3}+ 2\int
d_{12}S_{1,2}E^{+}_{1}E^{-}_{2}\nonumber\\
&&\hspace{0.4cm}+\int
d_{123}\Sigma_{1,2,3}(2E^{+}_{1}E^{-}_{2}E^{0}_{3}
+2E^{-}_{1}E^{+}_{2}E^{0}_{3}+2E^{0}_{1}E^{+}_{2}E^{-}_{3}),\label{q19}
\end{eqnarray}
where $E^{0}$ is the low-frequency virtual wave field and $E^{+},E^{-}$ are the high 
frequency fields and the frequencies of these fields are $\pm 1$.
The first two terms on the right hand side of Eq.~(\ref{q19}) describe
the nonlinearity of the $E^{0}$ field, (i.e. the low-frequency virtual wave field). 
This nonlinearity is negligible (in the first approximation) if $\varepsilon$ is not
small on the left-hand side of this equation. The last term on the
right hand side of Eq.~(\ref{q19}) produces the nonlinear contribution to
the dielectric permittivity of the low frequency field excited by
the high frequency fields ($E^{+},E^{-}$). If $\varepsilon$ is not small then this
term has the relative order of the ratio of the high frequency
energy to plasma particle energy. This ratio is the expansion
parameter allowing us to take into account only the first-order
nonlinearities. Therefore, under the mentioned condition ($\varepsilon$ is 
not close to zero), this term can
also be neglected in the first approximation. Thus we obtain~\cite{r4} 
\begin{eqnarray}
&&\hspace{-0.2cm}\varepsilon E^{0}\approx2\int
d_{23}S_{2,3}E^{+}_{2}E^{-}_{3},\label{q20}
\end{eqnarray}
where we have replaced the subscript indices $1$ and $2$ by $2$
and $3$, respectively. The equation for the $E^{+}$ field can be
written as
\begin{eqnarray}
&&\hspace{-0.2cm}\varepsilon E^{+}=2\int
d_{12}S_{1,2}E^{+}_{1}E^{0}_{2}+ 2\int
d_{12}S_{1,2}E^{-}_{1}E^{+2}_{2}\nonumber\\
&&\hspace{0.4cm}+2\int
d_{123}\Sigma_{1,2,3}E^{+}_{1}E^{+}_{2}E^{-}_{3} +\int
d_{123}\Sigma_{1,2,3}E^{-}_{1}E^{+}_{2}E^{+}_{3}.\label{q21}
\end{eqnarray}
The field on the second harmonic $E^{+2}$ is produced mainly by
the quadratic nonlinearity
\begin{eqnarray}
&&\hspace{-2.8cm}2=1+1,\label{q22}
\end{eqnarray}
and is determined by
\begin{eqnarray}
&&\hspace{-1.4cm}\varepsilon E^{+2}\approx\int
d_{23}S_{2,3}E^{+}_{2}E^{+}_{3}.\label{q23}
\end{eqnarray}
Substituting expressions Eqs.~(\ref{q20}) and (\ref{q23}) in Eq.~(\ref{q21}) we find
\begin{eqnarray}
&&\hspace{-1.8cm}\varepsilon E^{+}=2\int
d_{123}\Sigma^\mathrm{eff}_{1,2,3}E^{+}_{1}E^{+}_{2}E^{-}_{3} +\int
d_{123}\Sigma^\mathrm{eff}_{1,2,3}E^{-}_{1}E^{+}_{2}E^{+}_{3},\label{q24}
\end{eqnarray}
where
\begin{eqnarray}
&&\hspace{-2.8cm}\Sigma_{1,2,3}^\mathrm{eff}=\Sigma_{1,2,3}+\frac{2}{\varepsilon_{2+3}}S_{1,2+3}S_{2,3} \label{q25}
\end{eqnarray}
is the effective cubic plasma response, and $S$ and $\Sigma$ are
the symmetrized nonlinear responses defined by Eqs.~(\ref{q15})
and (\ref{q16}). The subscript $2+3$ denotes the dependence of the corresponding responses 
on ($\omega_2+\omega_3, {\bf k}_2+{\bf k}_3$), due to the variation driven by the fields 
2 and 3 combined, $\varepsilon_{2+3}$ is the total linear dielectric permittivity consists of 
the electron and ion contributions.

The first term on the right hand side of Eq.~(\ref{q24}) is much larger than the second one on the right hand side of Eq.~(\ref{q24}) (a detailed explanation is given in Appendix A). Finally the nonlinear equation for the high frequency field can be approximated as
\begin{eqnarray}
&&\hspace{-1.8cm}\varepsilon E^{+}\approx2\int
d_{123}\Sigma^\mathrm{eff}_{1,2,3}E^{+}_{1}E^{+}_{2}E^{-}_{3}.\label{q26}
\end{eqnarray}
The total linear dielectric permittivity of the plasma is 
\begin{eqnarray}
&&\hspace{-2.8cm}\varepsilon=\varepsilon^{(e)}+\varepsilon^{(i)}-1,\label{q27}
\end{eqnarray}
where $\varepsilon^{(e)}$ and $\varepsilon^{(i)}$ are
respectively the linear responses of the electron and ion
component of the quantum plasma. The minus 1 in the expression appears 
because the vacuum contribution was included both in the electron and ion dielectric 
permittivities.
The linear electron contribution can be expressed as
\begin{eqnarray}
&&\hspace{-1.8cm}\varepsilon^{(e)}_{2+3}=1-\frac{4\pi e^2}{m_e}\int
\frac{2d{\bf p}}{(2\pi\hbar)^3}
\frac{f_e^{(0)}(\bf p)}{(\omega_{2+3}-{\bf k}_{2+3}\cdot{\bf
v})^2-\left(\frac{\hbar k^2_{2+3}}{2m_e}\right)^2}.\label{q28}
\end{eqnarray}
We approximate the quadratic and cubic responses, which are contained
in $\Sigma^\mathrm{eff}_{1,2,3}$ [Eq.~(\ref{q25})] by making the
assumptions
\begin{eqnarray}
&&\hspace{-4.8cm} 1) \
\frac{\hbar k^2}{2m_e}\ll\omega_{pe},\nonumber\\
&&\hspace{-4.8cm} 2) ~\text{max} (kv_F, kv_{Te})\ll\omega_{pe},\nonumber\\
&&\hspace{-4.8cm}  3) ~\frac{\hbar {\bf k}\cdot {\bf
k}_1}{m_e}\lesssim \frac{\hbar k k_1}{m_e}\approx\frac{\hbar k^2}{m_e}\ll\omega_{pe} (\text{when} ~~ k_1\sim k),\nonumber\\
&&\hspace{-4.8cm} 4) ~|\Delta {\bf k}|=|{\bf k}-{\bf
k}_1|\lesssim|{\bf k}|.\label{q29}
\end{eqnarray}
We are considering the high-frequency field described by Eq.~(\ref{q26}).
For the quadratic response, we follow the assumptions $1)$ and $2)$
and finally obtain
\begin{eqnarray}
&&\hspace{-2.8cm}S_{1,2+3}=-\frac{1}{2}\frac{{\bf k}\cdot {\bf
k}_1}{|{\textbf{k}}||{\textbf{k}}_1|} \frac{|{\bf k}_2+{\bf k}_3|}{\omega^2_{pe}}\frac{i
e}{m_e}\left(\varepsilon^{(e)}_{2+3}-1\right),\label{q30}
\end{eqnarray}
and
\begin{eqnarray}
&&\hspace{-2.8cm}S_{2,3}=\frac{1}{2}\frac{{\bf k}_2\cdot\bf
{k_3}}{|{\textbf{k}}_2||{\textbf{k}}_3|} \frac{|{\bf k}_2+{\bf k}_3|}{\omega^2_{pe}}\frac{i
e}{m_e}\left(\varepsilon^{(e)}_{2+3}-1\right).\label{q31}
\end{eqnarray}
For the cubic response, we follow the assumptions $1)-4)$ and get
\begin{eqnarray}
&&\hspace{-2.8cm}\Sigma_{1,2,3}=\frac{1}{2}\frac{{\bf k}\cdot
{\bf k}_1}{|{\textbf{k}}||{\textbf{k}}_1|} \frac{{\textbf{k}}_2\cdot{\bf
k}_3}{|{\textbf k}_2||{\textbf k}_3|}\frac{e^2|{\bf k}_2+{\bf k}_3|^2}{m^2_e
\omega^2_{pe}}\left(\varepsilon^{(e)}_{2+3}-1\right).\label{q32}
\end{eqnarray}
The effective third-order response $\Sigma^\mathrm{eff}_{1,2,3}$  in Eq.~(\ref{q25}) can be
approximated in terms of the linear response as
\begin{eqnarray}
&&\hspace{-1.8cm}\Sigma^\mathrm{eff}_{1,2,3}=\frac{(1-\varepsilon^{(e)}_{2+3})\varepsilon^{(i)}_{2+3}}
{\varepsilon_{2+3}}
\frac{|{\bf k}_2+{\bf k}_3|^2}{8\pi n_{0}m_e\omega^2_{pe}}
\frac{({\bf k}_2\cdot{\bf k}_3)[{\bf k}_1\cdot({\bf k}_1+{\bf
k}_2+{\bf k}_3)]} {|{\bf k}_1||{\bf k}_2||{\bf k}_3||{\bf
k}_1+{\bf k}_2+{\bf k}_3|},\label{q33}
\end{eqnarray}
where the quantum plasma linear dielectric permittivities $\varepsilon_{2+3}^{(e)}$,  
$\varepsilon_{2+3}^{(i)}$ and $\varepsilon_{2+3}$ are given in the next section.
\section{The Effective Third Order Response in a Quantum Plasma}
The effective nonlinear response of a quantum plasma Eq.~(\ref{q33}) has the same form as 
the effective nonlinear response of a classical plasma, but
with the quantum-corrected linear responses, i.e., the quantum corrections only enter the
linear responses, not Eq.~(\ref{q33}) explicitly and this is our main result.
Below we consider separately two cases: non-degenerate and degenerate plasma 
electrons.
\subsection{Non-degenerate Plasma Electrons}
It is well known that a plasma is a many-particle system and for
its description it is thus natural to use methods of statistical
physics~\cite{r35,r36,r37}. Most often the plasma is in partial
thermodynamic equilibrium, and its components have different
temperatures with different equilibrium functions~\cite{r38,r39}.
The Maxwellian distribution of particles is possible only for
sufficiently high temperatures, when the Fermi degeneracy
following from the Pauli exclusion principle is negligible.  
When the thermal energy is large compared with the Fermi energy, 
the (Fermi-Dirac) distribution is well approximated by a Maxwellian  distribution, 
hence we use the Maxwellian distribution of electrons at equilibrium in this case. 
On the basis of the kinetic equation 
with self-consistent fields, the dielectric tensor of a homogeneous isotropic 
medium can be obtained by the equilibrium distribution function~\cite{r37}.
To derive the quantum-corrected Zakharov equations, we first derive 
the dielectric permittivity $\varepsilon$. The dispersion equation for 
longitudinal waves in any isotropic
plasma is $\varepsilon(\omega,{\bf k})=0$, where $\varepsilon$ is the
longitudinal dielectric function. For an electron-ion plasma in
the quantum case, it is suitable to define the quantum-corrected
electron ($\chi^{lq}_e$) and ion ($\chi^{lq}_i$) susceptibilities
such that the dielectric function becomes
\begin{eqnarray}
&&\hspace{-2.8cm}\varepsilon=1+\chi^{lq}_e(\omega,{\bf
k})+\chi^{lq}_i(\omega,{\bf k}).\label{q34}
\end{eqnarray}
The non-relativistic form for the longitudinal part of
the response tensor including the quantum recoil is~\cite{r19}
\begin{eqnarray}
&&\hspace{-2.8cm}\varepsilon=1-\sum_{\alpha}\frac{4\pi e^2}{m_{\alpha}}\int
d{\bf p}\frac{f_{\alpha}^{(0)}{(\bf p)}} {(\omega-{\bf k}\cdot{\bf
v})^2-\Delta_{\alpha}^2},\label{q35}
\end{eqnarray}
where $\alpha=e, i$. Here the quantum recoil for
electrons and ions is included through ${\Delta}_{\alpha}={\hbar k^2}/{2m_{\alpha}}$.
For a Maxwellian distribution of electrons and ions, the integral may be evaluated in
terms of the familiar plasma dispersion function. The singularity (i.e. 
${(\omega-kp_{\parallel}/m)^2-\Delta_{\alpha}^2}=0$ on the integration path over real $p_{\parallel}$) in the Eq.~(\ref{q35}) is avoided using Landau's rule~\cite{r40,r41,r42}, by replacing $\omega$ with $\omega+i0$ which in fact comes from the proper 
solution of the initial value problem using the Laplace transform in time. Landau's rule leads to complex-valued response function $\varepsilon(\omega,\mathbf{k})$, since we are ignoring the imaginary part of the response, it is sufficient to use the Fourier transform in time, and consider 
only the principal value part of the $p_x$ integral in Eq.~(\ref{q35}). The result is
\begin{eqnarray}
&&\hspace{-2.8cm}\int d{\bf p}\frac{f_{\alpha}^{(0)}({\bf p})} {\omega-{\bf
k}\cdot{\bf v}}=\frac{n_{\alpha}}{\sqrt{2}|{\bf k}|
v_{\alpha}} \frac{\phi(y_{\alpha})}{y_{\alpha}},
~~y_{\alpha}=\frac{\omega}{\sqrt{2}|{\bf k}|v_{\alpha}},\label{q36}
\end{eqnarray}
where the plasma dispersion function is defined by
\begin{eqnarray}
&&\hspace{-2.8cm}\phi(y)=-\frac{y}{\sqrt{\pi}}\int_{-\infty}^{\infty}\frac{
e^{-t^2}}{t-y}dt.\label{q37}
\end{eqnarray}
An alternative form for $\phi(y)$ for real $y$ is
\begin{eqnarray}
&&\hspace{-2.8cm}\phi(y)=2ye^{-y^2}\int_{0}^{y}e^{t^2}dt.\label{q38}
\end{eqnarray}
Expansion of the real part gives
\begin{eqnarray}
&&\hspace{-2.8cm}\phi(y)=y^2-\frac{4}{3}y^4+\cdot\cdot\cdot,
\textrm{for}~y^2\ll 1 ,\label{q39}
\end{eqnarray}
and
\begin{eqnarray}
&&\hspace{-2.8cm}\phi(y)=1+\frac{1}{2y^2}+\frac{1}{4y^4}+\cdot\cdot\cdot,
\textrm{for}~y^2\gg 1.\label{q40}
\end{eqnarray}
Equation~(\ref{q35}) is reduced to two terms of the form Eq.~(\ref{q36}) 
by writing
\begin{eqnarray}
&&\hspace{-2.8cm}\frac{1}{(\omega-{\bf k}\cdot{\bf
v})^2-\Delta_{\alpha}^2}=
\frac{1}{2{\Delta}_{\alpha}}\bigg(\frac{1}{\omega-{\bf k}\cdot{\bf
v}-{\Delta}_{\alpha}}-\frac{1}{\omega-{\bf k}\cdot{\bf
v}+{\Delta}_{\alpha}}\bigg).\label{q41}
\end{eqnarray}
Then Eq.~(\ref{q36}) implies that
\begin{eqnarray}
&&\hspace{-2.8cm}\int d{\bf p}\frac{f_{\alpha}^{(0)}({\bf p})} {\omega-{\bf
k}\cdot{\bf v}\pm{\Delta}_{\alpha}}=\frac{n_{\alpha}}{\sqrt{2}|{\bf k}|
v_{\alpha}} \frac{\phi(y_{\pm\alpha})}{y_{\pm\alpha}}, \label{q42}
\end{eqnarray}
with
\begin{eqnarray}
&&\hspace{-2.8cm}{y_{\pm\alpha}}=\frac{\omega\pm{\Delta}_{\alpha}}{\sqrt{2}|{\bf k}|
v_{\alpha}}. \label{q43}
\end{eqnarray}
Then the electron and ion susceptibilities of a Maxwellian plasma, with the quantum recoil included, are~\cite{r19}

\begin{eqnarray}
&&\hspace{-1.8cm}\chi^{lq}_{\alpha}(\omega,{\bf
k})=-\frac{\omega_{p\alpha}^2}{\sqrt{2}|{\bf k}|
v_{\alpha}}\frac{1}{2{\Delta}_{\alpha}}\left[\frac{\phi(y_{-\alpha})}
{y_{-\alpha}}-\frac{\phi(y_{+\alpha})}{y_{+\alpha}}\right].\label{q44}
\end{eqnarray}
For high frequency Langmuir waves Eq.~(\ref{q44}) is approximated under the
assumption that the phase speed is greater than the electron
thermal speed (i.e. $\omega\gg kv_{Te}$). Using the approximation
 $y_{\pm e}\gg 1$ the electron susceptibility becomes
\begin{eqnarray}
&&\hspace{-2.8cm}\chi^{lq}_{e}(\omega,{\bf
k})=-\frac{\omega^2_{pe}}{\omega^2-\Delta^2_e}\left[1+\frac{3k^2v^2_{Te}}{\omega^2
}\right].\label{q45}
\end{eqnarray}
The approximation for the high frequency linear dielectric
permittivity is then
\begin{eqnarray}
&&\hspace{-2.8cm}\varepsilon^{(e)}=1-\frac{\omega^2_{pe}}{\omega^2}-3\frac{k^2v^2_{Te}}
{\omega^2_{pe}}-\frac{\hbar^2k^4}{4m^2_e\omega^2_{pe}}.\label{q46}
\end{eqnarray}
For low frequency plasma density perturbation Eq.~(\ref{q44}) is approximated 
under the assumption that the phase speed is intermediate
between the electron and the ion thermal speeds (i.e. $kv_{Ti}\ll
\omega\ll kv_{Te}$). In a fluid approach, this corresponds to the
electrons behaving isothermally and the ions behaving
adiabatically; in a kinetic approach this corresponds to making
the approximations $y_{\pm e}\ll 1$ and $y_{\pm i}\gg 1$. Using
the approximation $y_{\pm e}\ll 1$ the electron susceptibility
becomes
\begin{eqnarray}
&&\hspace{-2.8cm}\chi^{lq}_{e}(\omega,{\bf k})=\frac{1}{k^2\lambda^2_{De}}\left[1-\frac{(3\omega^2+\Delta^2_e)}{3k^2v^2_{Te}
}\right].\label{q47}
\end{eqnarray}
Using the approximation $y_{\pm i}\gg 1$ and neglecting the
quantum recoil term the ion susceptibility becomes
\begin{eqnarray}
&&\hspace{-4.8cm}\chi^{lq}_{i}(\omega,{\bf
k})=-\frac{\omega^2_{pi}}{\omega^2}.\label{q48}
\end{eqnarray}
The approximation for low frequency linear dielectric
permittivity of the Maxwellian plasma with the electron quantum recoil taken into account is then
\begin{eqnarray}
&&\hspace{-2.8cm}\varepsilon=1-\frac{\omega^2_{pi}}{\omega^2}+\frac{\omega^2_{pe}}
{k^2v^2_{Te}}\left(1-\frac{\hbar^2k^2}{12m^2_ev^2_{Te}}\right).\label{q49}
\end{eqnarray}
\subsection{Fully Degenerate Plasma electrons $(T_e=0)$} It is
well known that the degeneracy of a Fermi system, i.e. particles
with half-integer spin, becomes important when the Fermi energy becomes 
comparable to, or exceeds the thermal energy~\cite{r37}. Here we consider the case of 
fully degenerate electrons, and derive the linear permittivity of 
such plasma.

The ratios of the energy spread of ions (defined by $v_{Ti}$ for non-degenerate ions,
or $v_{Fi}$ for degenerate ions) to the energy spread of electrons 
(defined by $v_{Fe}$) $v_{Fi}/v_{Fe}$ (i.e. $\sqrt{(m_e/m_i)(T_{Fi}/T_{Fe})}$) 
for degenerate ions and for non-degenerate ions $v_{Ti}/v_{Fe}$ (i.e. $\sqrt{(m_e/m_i)(T_i/T_{Fe})}$) 
are small. Because the ratio of the  electron and ion mass ($m_e/m_i$) is small 
and the ratio of the ion and electron temperature ($T_i/T_e$) is usually small and ($T_{Fi}/T_{Fe}$), ($T_i/T_{Fe}$) are also small. Since ions are much heavier than electrons, the ion susceptibility rather comes from neglecting the energy spread of ions  which is small compared to the energy spread of
electrons.

Substituting $\tilde{f}_e^{(1)}$ in Poisson's equation with the help of Eq.~(\ref{q8}), we obtain the electron susceptibility
\begin{eqnarray}
&&\hspace{-1.8cm}\chi^{lq}_e=\frac{4\pi e^2
}{k^2\hbar}\int\frac{1}{(\omega-{\bf k} \cdot{\bf v})}\left[f_e^{(0)}\left({\bf v}+\frac{\hbar {\bf
k}}{2m_e}\right)-f_e^{(0)}\left({\bf v}-\frac{\hbar{\bf
k}}{2m_e}\right)\right]d{\bf v}.\label{q50}
\end{eqnarray}
After changing variables Eq.~(\ref{q50}) becomes
\begin{eqnarray}
&&\hspace{-1.8cm}\chi^{lq}_e=\frac{4\pi e^2
}{k^2\hbar}\int\left[\frac{1}{\left[\omega-{\bf k}
\cdot\left({\bf u}-\frac{\hbar{\bf k}}{2m_e}\right)\right]}
-\frac{1}{\left[\omega-{\bf k} \cdot\left({\bf
u}+\frac{\hbar{\bf k}}{2m_e}\right)\right]}\right]f_e^{(0)}({\bf
u})d{\bf u},\label{q51}
\end{eqnarray}
which can be written as
\begin{eqnarray}
&&\hspace{-2.8cm}\chi^{lq}_e=-\frac{4\pi
e^2}{m_e}\int\frac{f_e^{(0)}(\bf u)}{{(\omega-{\bf k} \cdot{\bf u})^2}
-\frac{\hbar^2k^4}{4m^2_e}}d{\bf u}.\label{q52}
\end{eqnarray}
The electron susceptibility was also derived by Bohm and Pines
\cite{r43} using a series of canonical transformations of the
Hamiltonian of the system. Now, we choose a coordinate system such that the $x$ axis
is aligned with the wave vector $\bf k$. Then Eq.~(\ref{q52}) takes
the form
\begin{eqnarray}
&&\hspace{-2.8cm}\chi^{lq}_e=-\frac{4\pi
e^2}{m_e}\int\frac{f_e^{(0)}(\bf u)}{{(\omega-k u_x)^2}
-\frac{\hbar^2k^4}{4m^2_e}}d{\bf u}.\label{q53}
\end{eqnarray}
We consider a plasma with degenerate electrons in the zero-temperature limit, (e.g., 
$T_e<<E_{Fe}$). Then, the background distribution function takes the simple form
\begin{align}
&&\hspace{-7.6cm}f_e^{(0)}({\bf
u}) = \begin{cases} 2\left(\frac{m_e}{2\pi
\hbar}\right)^3, &|{\bf u}|\leq V_{Fe}, \\
0, & \textrm{elsewhere}, \end{cases} ~~~~~~~~~~~\label{q54}
\end{align}
where $V_{Fe}$ is the speed of an electron on the Fermi surface.
The integration can be performed over velocity space
perpendicular to $u_x$, using cartesian coordinates $u_y$ and
$u_z$, and thus Eq.~(\ref{q53}) can be written as
\begin{eqnarray}
&&\hspace{-2.8cm}\chi^{lq}_e=-\frac{4\pi
e^2}{m_e}\int\frac{F_e^{(0)}({u_x})}{{(\omega-k u_x)^2}
-\frac{\hbar^2k^4}{4m^2_e}}d{u_x},\label{q55}
\end{eqnarray}
where 
\begin{eqnarray}
&&\hspace{-2.8cm}F_e^{(0)}({u_x})=\iint f_e^{(0)}({\bf u})du_ydu_z=2\pi\int_0^{\sqrt{V^2_{Fe}-u^2_x}}2\left(\frac{m_e}{2\pi
\hbar}\right)^3u_{\bot}du_{\bot},\label{q56}
\end{eqnarray}
\begin{align}
&&\hspace{-5.0cm} = \begin{cases} 2\pi\left(\frac{m_e}{2\pi
\hbar}\right)^3(V^2_{Fe}-u^2_x), &|{u_x}|\leq V_{Fe}, \\
0, & \textrm{elsewhere}. \end{cases} ~~~~~~~~~~~~~~~~~~\label{q57}
\end{align}
Then Eq.~(\ref{q55}) can be written as
\begin{eqnarray}
&&\hspace{-2.8cm}\chi^{lq}_e=-\frac{3
\omega^2_{pe}}{4V^3_{Fe}}\int_{-V_{Fe}}^{V_{Fe}}\frac{V^2_{Fe}-u^2_x}
{(\omega-ku_x)^2-\frac{\hbar^2k^4}{4m^2_e}}du_x.\label{q58}
\end{eqnarray}
Performing the integration over $u_x$ (using standard Landau's rule)~\cite{r40,r41,r42}, 
we get from Eq.~(\ref{q58})~\cite{r44}
\begin{eqnarray}
&&\hspace{-1.8cm}\chi^{lq}_e=\frac{3
\omega^2_{pe}}{4k^2V^2_{Fe}}\bigg\{2-\frac{m_e}{\hbar
kV_{Fe}}\bigg[V^2_{Fe}-\bigg(\frac{\omega}{k}+\frac{\hbar k
}{2m_e}\bigg)^2\bigg]\ln\bigg(
\frac{\omega-kV_{Fe}+\frac{\hbar k^2
}{2m_e}}{\omega+kV_{Fe}+\frac{\hbar k^2
}{2m_e}}
\bigg)\nonumber\\
&&\hspace{-0.8cm}~~~~~+\frac{m_e}{\hbar
kV_{Fe}}\bigg[V^2_{Fe}-\bigg(\frac{\omega}{k}-\frac{\hbar k
}{2m_e}\bigg)^2\bigg]\ln\bigg(
\frac{\omega-kV_{Fe}-\frac{\hbar k^2
}{2m_e}}{\omega+kV_{Fe}-\frac{\hbar k^2
}{2m_e}}
\bigg)\bigg\},\label{q59}
\end{eqnarray}
where principle branch of the complex log function defined as
\begin{align}
&&\hspace{-7.6cm}\textrm{ln}z = \begin{cases} \textrm{ln}|z|-i\pi, &z<0, \\
\textrm{ln}z,&z\geq 0. \end{cases} ~~~~~~~~~~~\label{q60}
\end{align}
Since we are ignoring the imaginary part of the response 
function $\varepsilon(\omega,\mathbf{k})$, it is sufficient to use the Fourier transform in time, and consider only the principal value part of the $u_x$ integral in Eq.~(\ref{q59}). The result is
\begin{eqnarray}
&&\hspace{-1.8cm}\chi^{lq}_e=\frac{3
\omega^2_{pe}}{4k^2V^2_{Fe}}\bigg\{2-\frac{m_e}{\hbar
kV_{Fe}}\bigg[V^2_{Fe}-\bigg(\frac{\omega}{k}+\frac{\hbar k
}{2m_e}\bigg)^2\bigg]\ln\bigg|
\frac{\omega-kV_{Fe}+\frac{\hbar k^2
}{2m_e}}{\omega+kV_{Fe}+\frac{\hbar k^2
}{2m_e}}
\bigg|\nonumber\\
&&\hspace{-0.8cm}~~~~~+\frac{m_e}{\hbar
kV_{Fe}}\bigg[V^2_{Fe}-\bigg(\frac{\omega}{k}-\frac{\hbar k
}{2m_e}\bigg)^2\bigg]\ln\bigg|
\frac{\omega-kV_{Fe}-\frac{\hbar k^2
}{2m_e}}{\omega+kV_{Fe}-\frac{\hbar k^2
}{2m_e}}
\bigg|\bigg\}.\label{q61}
\end{eqnarray}
For a high frequency wave (i.e. $\omega\gg
kV_{Fe}\gg\frac{\hbar k^2}{2m_e}$) the approximation for the high
frequency linear dielectric permittivity of degenerate plasma is
\begin{eqnarray}
&&\hspace{-2.8cm}\varepsilon^{(e)}=1-\frac{\omega^2_{pe}}{\omega^2}-\frac{3}{5}
\frac{k^2V^2_{Fe}}{\omega^2_{pe}}-\frac{\hbar^2k^4}{4m^2_e\omega^2_{pe}}.\label{q62}
\end{eqnarray}
In a quantum plasma system consisting of mobile ions and
inertialess electrons, we have the possibility of low frequency
waves (in comparison with the electron plasma frequency)
in which case the approximation for low frequency (i.e. $\omega\ll \omega_{pe},kV_{Fe}$)
linear dielectric permittivity is
\begin{eqnarray}
&&\hspace{-2.8cm}\varepsilon=1-\frac{\omega^2_{pi}}{\omega^2}+\frac{3\omega^2_{pe}}{k^2V^2_{Fe}}
\left(1-\frac{\hbar^2k^2}{12m^2_ev^2_{Fe}}\right).\label{q63}
\end{eqnarray}
\section{Kinetic Quantum Zakharov Equations}
In order to obtain the set of equations describing the
nonlinear interaction between high frequency Langmuir waves and low-frequency 
plasma density variations, in the quantum regime, we follow the 
kinetic derivation of the Zakharov equations in the classical case 
that was carried out by Vladimirov et al.\cite{r4}. A general discussion
of the validity of the Zakharov equations can be found in the
review paper by Thornhill and ter Haar~\cite{r6}.

The approximation Eq.~(\ref{q33}) is the main result of this
exercise; below we use it to derive the Zakharov equations for
a quantum plasma. The low-frequency plasma density
variations are derived in Appendix B and are given by
\begin{eqnarray}
&&\hspace{-2.8cm}\frac{\delta
n}{n_0}=-\frac{\varepsilon^{(i)}}{\varepsilon}\left(1-\frac{\hbar^2
k^2}{12m^2_e v^2_e}\right)\int \frac{E^{+}_2 E^{-}_3}{4\pi
n_0T_e} \frac{({\bf k}_2\cdot{\bf k}_3)}{|{\bf k}_2||
{\bf k}_3|}d_{23}. \label{q64}
\end{eqnarray}
To find the dynamical evolution equation for the slowly varying amplitude of
the high-frequency field $E({\bf r},t)$, we use the inverse Fourier transform.

The Fourier transform of the fast oscillating total electric field is
\begin{eqnarray}
&&\hspace{-2.8cm}E({\bf r},t)=\frac{1}{(2\pi)^4}\int
d\omega d{\bf k} \frac{{\bf k}}{|{\bf k} |} E^{+}_{\omega,{\bf
k}} e^{i({\bf k}\cdot{\bf r}-\omega t)}.\label{q65}
\end{eqnarray}
The slowly-varying envelope of the Langmuir wave electric field
can be written as
\begin{eqnarray}
&&\hspace{-1.2cm}E_{env}({\bf r},t)\simeq E({\bf r},t)
e^{i\omega_{pe}t}\simeq\frac{1}{(2\pi)^4}\int d\omega d{\bf k}
\frac{{\bf k}}{|{\bf k}|}E^{+}_{\omega,{\bf k}} e^{i({\bf
k}\cdot{\bf r}-\omega t+\omega_{pe} t)}. \label{q66}
\end{eqnarray}
The high- and low-frequency approximations for the linear
response functions for both non-degenerate and fully degenerate 
plasma electrons are $\varepsilon^{(e)}$ [Eqs.~(\ref{q46}), (\ref{q62})] 
and $\varepsilon$ [Eqs.~(\ref{q49}), (\ref{q63})].
Using the values of Eqs.~(\ref{q64}), (\ref{q66})
and $\varepsilon^{(e)}$ in Eq.~(\ref{q26}), in the one-dimensional 
case we obtain from Eq.~(\ref{q26}) the equation for the evolution 
of high frequency field in the form
\begin{eqnarray}
&&\hspace{-1.8cm}\left(i \frac{\partial}{\partial
t}+P\frac{v^2_e}{\omega_{pe}}\frac{\partial^2}{\partial x^2}
-\frac{\hbar^2}{8m^2_e\omega_{pe}}\frac{\partial^4}{\partial
x^4}\right)E(x,t) =\frac{\omega_{pe}}{2}\frac{\delta n
(x,t)}{n_0}E(x,t),\label{q67}
\end{eqnarray}
Furthermore, using the definition of plasma density variation
in Eq.~(\ref{q64}), and using Eqs.~(\ref{q49}), (\ref{q63}) and 
the value of $\varepsilon^{(i)}$, 
we get the equation for low frequency evolution of
plasma density $\delta n$ in the following form:
\begin{eqnarray}
&&\hspace{-1.8cm}\left(\frac{\partial^2}{\partial
t^2}-Qv^2_s\frac{\partial^2}{\partial x^2}+\frac{\hbar^2}{12m^2_ev^2_e}\frac{\partial^4}{\partial t^2
\partial x^2}
\right)\frac{\delta n
(x,t) }{n_0} =\left[\frac{\partial^2}{\partial
x^2}+\frac{\hbar^2}{12m^2_ev^2_e}\frac{\partial^4}{\partial x^4}
\right]\frac{|E(x,t)|^2}{4\pi n_0 m_i}, \label{q68}
\end{eqnarray}
where $P=3/2,\ Q=1$ for non-degenerate plasma electrons,
and $P=3/10,\ Q=1/3$ for fully degenerate plasma electrons. The
resulting quantum-corrected Zakharov equations
in Eqs.~(\ref{q67}) and (\ref{q68}) describe the coupled nonlinear evolution
of high-frequency fields and low-frequency density variations in
collisionless quantum plasmas.
\section{Discussion and conclusion}
In this work, we have presented the formalism of
modulational interactions in quantum plasmas based on the Wigner
kinetic description of collisionless quantum plasmas. 
Under assumptions 1)-4), we have shown that the nonlinear response of the plasma 
to electrostatic fields can be described by the effective cubic response function, 
which takes into account $2$- and $3$-wave interactions in the quantum plasma.
We stress that the effective cubic response 
of a quantum plasma has the same form as the effective 
cubic response of a classical plasma, but with the quantum-corrected linear responses, 
i.e., the quantum corrections only enter the linear responses, not Eq.~(\ref{q33}) 
explicitly; and this is our new result. The derived Zakharov equations for a quantum plasma are 
an illustration of how the formalism can be applied. The effective cubic response 
has both real and imaginary parts, while the Zakharov equations only account for the 
real part of the effective response. The effective cubic response has more information 
and can also be used for other problems; in particular, for the calculation of 
nonlinear Landau damping of coherent structures.
We have used the assumptions $k\lambda_{De}\ll1$ ($\lambda_{De}$ is the Debye length for non-degenerate 
plasma electrons) and $k\lambda_{F}\ll1$ ($\lambda_{F}$ is the Debye length for degenerate 
plasma electrons) in our kinetic derivation of the effective cubic response ($\Sigma^\mathrm
{eff}_{1,2,3}$). Hence our kinetic theory is limited to $k\lambda_{De}\ll1$ and $k\lambda_{F}\ll1$ 
which is the same region of validity as quantum fluid theory (QFT). In the fluid theory the procedure 
of averaging over the high frequency (the electron plasma frequency $\omega_{pe}$) has been used to derive the Zakharov equations~\cite{r31}. 
This procedure does not allow a proper treatment of the so-called higher and electron nonlinearities straightforwardly and this is the applicability limit in the fluid theory. The kinetic derivation is more general and thus more rigorous, and allows to establish the limits of applicability of the results, which are not at all clear within the fluid theory~\cite{r21,r31}. Moreover, our Zakharov equations do not match those derived from QFT~\cite{r21} exactly. In the left hand side Eq.~(\ref{q68}), the quantum  correction term [${\hbar^2}/{12m_e^2v_e^2}({\partial^4/\partial x^2\partial t^2})$] is different from the corresponding term [${\hbar^2}/{4m_im_e}({\partial^4/\partial x^4})$] obtain in the framework of QFT. Because in the QFT~\cite{r21} a real low-frequency wave is assumed, in our  case we assume a virtual low-frequency wave instead of a real low-frequency wave. The right-hand side of Eq.~(\ref{q68}) for the low frequency evolution of plasma density variations also has an extra term compared to the corresponding equation of QFT~\cite{r21}. This correction term comes from the  dynamics of the quantum electrons that is taken from the kinetic  theory which changes the ponderomotive force by a small quantum correction  term proportional to $\hbar^2$. The general form of the set of equations (Zakharov equations) for both non-degenerate and fully degenerate plasma electrons is the same. Only the coefficients ($P$ and $Q$) are different for non-degenerate and fully degenerate plasma electrons. 

The modulational interactions are 
significant for highly non-equilibrium systems. 
The presented formalism can be used in a range of problems including 
development of turbulence, the process of self-organization, 
as well as transition from weak turbulent state to strong turbulence.
These phenomena result in the formation of strongly correlated
structures (solitons, cavitons, etc.), generation of strong
magnetic fields, heating, and effective particle 
acceleration~\cite{r1,r2,r3,r4,r5,r6,r7,r8,r9,r10}.
\appendix
\section{}
The equation for the $E^{+}$ field can be
written as
\begin{eqnarray}
&&\hspace{-1.8cm}\varepsilon E^{+}=2\int
d_{123}\Sigma^\mathrm{eff}_{1,2,3}E^{+}_{1}E^{+}_{2}E^{-}_{3} +\int
d_{123}\Sigma^\mathrm{eff}_{1,2,3}E^{-}_{1}E^{+}_{2}E^{+}_{3},\label{a1}
\end{eqnarray}
where
\begin{eqnarray}
&&\hspace{-2.8cm}\Sigma_{1,2,3}^\mathrm{eff}=\Sigma_{1,2,3}+\frac{2}{\varepsilon_{2+3}}S_{1,2+3}S_{2,3} \label{a2}
\end{eqnarray}
is the effective cubic plasma response, and $S$ and $\Sigma$ are
the symmetrized nonlinear responses defined by Eqs.~(\ref{q15})
and (\ref{q16}). The subscript $2+3$ denotes the dependence of the corresponding responses 
on ($\omega_2+\omega_3, {\bf k}_2+{\bf k}_3$), due to the variation driven by the fields 
$2$ and $3$ combined. Under assumptions 1)-4), one can approximate the quadratic and 
cubic responses in Eq.~(\ref{a2}) by
\begin{eqnarray}
&&\hspace{-2.8cm}S_{1,2+3}=-\frac{1}{2}\frac{{\bf k}\cdot {\bf
k}_1}{|{\textbf{k}}||{\textbf{k}}_1|} \frac{|{\bf k}_2+{\bf k}_3|}{\omega^2_{pe}}\frac{i
e}{m_e}\left(\varepsilon^{(e)}_{2+3}-1\right),\label{a3}
\end{eqnarray}
and
\begin{eqnarray}
&&\hspace{-2.8cm}S_{2,3}=\frac{1}{2}\frac{{\bf k}_2\cdot\bf
{k_3}}{|{\textbf{k}}_2||{\textbf{k}}_3|} \frac{|{\bf k}_2+{\bf k}_3|}{\omega^2_{pe}}\frac{i
e}{m_e}\left(\varepsilon^{(e)}_{2+3}-1\right).\label{a4}
\end{eqnarray}
For the cubic response, we get
\begin{eqnarray}
&&\hspace{-2.8cm}\Sigma_{1,2,3}=\frac{1}{2}\frac{{\bf k}\cdot
{\bf k}_1}{|{\textbf{k}}||{\textbf{k}}_1|} \frac{{\textbf{k}}_2\cdot{\bf
k}_3}{|{\textbf k}_2||{\textbf k}_3|}\frac{e^2|{\bf k}_2+{\bf k}_3|^2}{m^2_e
\omega^2_{pe}}\left(\varepsilon^{(e)}_{2+3}-1\right),\label{a5}
\end{eqnarray}
where
\begin{eqnarray}
&&\hspace{-1.8cm}\varepsilon^{(e)}_{2+3}-1=-\frac{4\pi e^2}{m_e}\int
\frac{2d{\bf p}}{(2\pi\hbar)^3}
\frac{f_e^{(0)}(\bf p)}{(\omega_{2+3}-{\bf k}_{2+3}\cdot{\bf
v})^2-\bigg(\frac{\hbar k^2_{2+3}}{2m_e}\bigg)^2}.\label{a6}
\end{eqnarray}
Thus the first term on the right hand side of Eq.~(\ref{a1}) contains the multiplier
\begin{eqnarray}
&&\hspace{-1.8cm}\varepsilon^{(e)}_{\omega_2^{+}+\omega_3^{-},{\bf k}_2^{+}+{\bf k}_3^{-}}-1\nonumber\\
&&\hspace{-1.8cm}=-\frac{4\pi e^2}{m_e}\int
\frac{2d{\bf p}}{(2\pi\hbar)^3}
\frac{f_e^{(0)}(\bf p)}{[(\omega_2^{+}+\omega_3^{-})-({\bf k}_2^{+}+{\bf k}_3^{-})\cdot{\bf
v}]^2-\bigg[\frac{\hbar({\bf k}_2^{+}+{\bf k}_3^{-})^2}{2m_e}\bigg]^2},~~~~\label{a7}
\end{eqnarray}
where the difference in the frequencies of the interacting waves, $\omega - \omega_1^{+}=\omega_2^{+} + \omega_3^{-} \ll \omega_{pe}$.
The denominator of Eq.~(\ref{a7}) can be written as
\begin{eqnarray}
&&\hspace{-1.8cm}{[(\omega_2^{+}+\omega_3^{-})-({\bf k}_2^{+}+{\bf k}_3^{-})\cdot{\bf
v}]^2-\bigg[\frac{\hbar({\bf k}_2^{+}+{\bf k}_3^{-})^2}{2m_e}}\bigg]^2\nonumber\\
&&\hspace{-1.8cm}=\bigg[\bigg(\omega_2^{+}-{\bf k}_2^{+}\cdot{\bf
v}+\frac{\hbar k^2_2}{2m_e}\bigg)+\bigg(\omega_3^{-}-{\bf k}_3^{-}\cdot{\bf
v}+\frac{\hbar k^2_3}{2m_e}+\bigg)+\frac{\hbar {\bf k}_2^{+}\cdot {\bf
k}_3^{-}}{m_e}\bigg]\nonumber\\
&&\hspace{-1.8cm}\bigg[\bigg(\omega_2^{+}-{\bf k}_2^{-}\cdot{\bf
v}-\frac{\hbar k^2_2}{2m_e})+\bigg(\omega_3^{-}-{\bf k}_3^{-}\cdot{\bf
v}-\frac{\hbar k^2_3}{2m_e}\bigg)-\frac{\hbar {\bf k}_2^{+}\cdot {\bf
k}_3^{-}}{m_e}\bigg].\label{a8}
\end{eqnarray}
When $k_j\sim k$, where $j=2$ and $3$, Eq.~(\ref{a8}) can be approximated under the following assumptions:
\begin{eqnarray}
&&\hspace{-4.8cm} 1) \
\frac{\hbar k^2}{2m_e}\ll \omega_{pe},\nonumber\\
&&\hspace{-4.8cm} 2) ~\text{max}(kv_F, kv_{Te})\ll \omega_{pe},\nonumber\\
&&\hspace{-4.8cm}  3) \frac{\hbar {\bf k}_2\cdot {\bf
k}_3}{m_e}\lesssim \frac{\hbar k_2 k_3}{m_e}\approx\frac{\hbar k^2}{m_e}\ll\omega_{pe}.\label{a9}
\end{eqnarray}
Under the assumptions 1)-3) the denominator of Eq.~(\ref{a7}) is approximately $(\omega^{+}+\omega^{-})^2<<\omega_{pe}^2$ (since $\omega^{-}\sim -\omega^{+}$).
\\\\
The second term on the right hand side of Eq.~(\ref{a1}) contains the multiplier
\begin{eqnarray}
&&\hspace{-1.8cm}\varepsilon^{(e)}_{\omega_2^{+}+\omega_3^{+},{\bf k}_2^{+}+{\bf k}_3^{+}}-1\nonumber\\
&&\hspace{-1.8cm}=-\frac{4\pi e^2}{m_e}\int
\frac{2d{\bf p}}{(2\pi\hbar)^3}
\frac{f_e^{(0)}(\bf p)}{[(\omega_2^{+}+\omega_3^{+})-({\bf k}_2^{+}+{\bf k}_3^{+})\cdot{\bf
v}]^2-\bigg[\frac{\hbar({\bf k}_2^{+}+{\bf k}_3^{+})^2}{2m_e}\bigg]^2},~~~~\label{a10}
\end{eqnarray}
where the difference in the frequencies of the interacting waves, $\omega - \omega_1^{-}=\omega_2^{+} + \omega_3^{+}\sim 2\omega_{pe}$ (since $\omega_{2,3}^{+}\sim\omega_{pe}$).
\\\\
Similarly to the denominator of Eq.~(\ref{a7}), the denominator of Eq.~(\ref{a10}) is approximately $(\omega^{+}+\omega^{+})^2\sim4\omega_{pe}^2$.
\\\\
The first term on the right hand side of Eq.~(\ref{a1}) is much larger than the second one on the right hand side of Eq.~(\ref{a1}) because of it's small denomitor (i.e. $\omega^{+} + \omega^{-} \ll \omega_{pe}$). So, the nonlinearity due to the second term of Eq.~(\ref{a1}) is negligible. Finally the nonlinear equation for the high frequency field can be approximated as
\begin{eqnarray}
&&\hspace{-1.8cm}\varepsilon E^{+}\approx2\int
d_{123}\Sigma^\mathrm{eff}_{1,2,3}E^{+}_{1}E^{+}_{2}E^{-}_{3}.\label{b11}
\end{eqnarray}
\section{}
In this appendix we are deriving the defination of the low-frequency plasma density
variations ($\delta n/{n_0}$).
The approximation for the high frequency field equation can be
written as
\begin{eqnarray}
&&\hspace{-2.8cm}\varepsilon E^{+}\approx2\int
d_{123}\Sigma^\mathrm{eff}_{1,2,3}E^{+}_1E^{+}_2E^{-}_3,\label{b1}
\end{eqnarray}
where the effective cubic response $\Sigma^\mathrm{eff}_{1,2,3}$ can be
written as
\begin{eqnarray}
&&\hspace{-2.8cm}\Sigma^\mathrm{eff}_{1,2,3}=\frac{(1-\varepsilon^{(e)}_{2+3})
\varepsilon^{(i)}_{2+3}}{\varepsilon_{2+3}}
\frac{\mid {\bf k}_2+{\bf k}_3 \mid^2}{8\pi n_{0}m_e\omega^2_{pe}}
\frac{({\bf k}_2\cdot{\bf k}_3)[{\bf k}_1\cdot({\bf k}_1+{\bf
k}_2+{\bf k}_3)]} {|{\bf k}_1||{\bf k}_2||
{\bf k}_1+{\bf k}_2+{\bf k}_3|},\label{b2}
\end{eqnarray}
and the right hand side of Eq.~(\ref{b1}) gives the value of ($\delta n/{n_0}$).
Now the right hand side of Eq.~(\ref{b1}) is
\begin{eqnarray}
&&\hspace{-1.4cm}2\int
d_{123}\Sigma^\mathrm{eff}_{1,2,3}E^{+}_1E^{+}_2E^{-}_3=-\int
d_{123}\frac{\varepsilon^{(i)}}{\varepsilon} \frac{1}{4\pi
n_{0}m_e\omega^2_{pe}\lambda^2_{De}}\left(1-\frac{\hbar^2
k^2}{12m^2_ev^2_e}\right)\nonumber\\
&&\hspace{2.8cm}\frac{[{\bf k}_1\cdot({\bf k}_1+{\bf k}_2+{\bf
k}_3)]} {|{\bf k}_1||{\bf k}_1+{\bf k}_2+{\bf
k}_3|} E^{+}_1\frac{{\bf k}_2\cdot{\bf k}_3}{|{\bf
k}_2||{\bf k}_3|}E^{+}_2E^{-}_3 \label{b3}
\end{eqnarray}
\begin{eqnarray}
&&\hspace{0.4cm}=-\int\frac{1}{4\pi n_{0}T_e} \left(1-\frac{\hbar^2
k^2}{12m^2_ev^2_e}\right) \frac{{\bf k}\cdot{\bf k}_1}
{kk_1}E^{+}_1\left(\frac{\varepsilon^{(i)}}{\varepsilon}
\frac{{\bf k}_2\cdot{\bf k}_3}{|{\bf k}_2||{\bf
k}_3|}E^{+}_2E^{-}_3\right)d_{123} \label{b4}
\end{eqnarray}
\begin{eqnarray}
&&\hspace{0.2cm}=-\int \frac{d1~d\varDelta~\delta(\omega-\omega_1-\varDelta\omega)\delta({\bf k}-{\bf
k}_1-\varDelta{\bf k})}{4\pi n_{0}T_e} \left(1-\frac{\hbar^2
k^2}{12m^2_ev^2_e}\right) \frac{{\bf k}\cdot{\bf k}_1}
{kk_1}E^{+}_1\nonumber\\
&&\hspace{1.6cm}\bigg\{\frac{\varepsilon^{(i)}}{\varepsilon}
\frac{{\bf k}_2\cdot{\bf k}_3}{|{\bf k}_2||\bf
{k_3}|}E^{+}_2E^{-}_3
d2~d3~\delta(\varDelta\omega-\omega_2-\omega_3)\delta(\varDelta{\bf k}-{\bf k}_2-{\bf
k}_3)\bigg\} \label{b5}
\end{eqnarray}
\begin{eqnarray}
&&\hspace{-0.8cm}=\int \frac {d_{1\varDelta}}{4\pi n_{0}T_e} \left(1-\frac{\hbar^2
k^2}{12m^2_ev^2_e}\right)\frac{{\bf k}\cdot{\bf k}_1}
{kk_1}E^{+}_1\frac{\varepsilon^{(i)}}{\varepsilon}
\int\frac{{\bf k}_2\cdot{\bf k}_3}{|{\bf k}_2||\bf
{k_3}|}E^{+}_2E^{-}_3 d_{23} \label{b6}
\end{eqnarray}
\begin{eqnarray}
&&\hspace{-8.2cm}=\int d_{1\varDelta} \frac{{\bf k}\cdot{\bf k}_1}
{kk_1}E^{+}_1\frac{\delta n}{n_0}, \label{b7}
\end{eqnarray}
where 

$d_{123}=d\omega_1 ~d{\bf k}_1 ~d\omega_2
~d{\bf k}_2 ~d\omega_3 ~d{\bf k}_3 ~\delta({\bf k}-{\bf k_{1}}-{\bf k_{2}}-{\bf
k_{3}})\delta(\omega-\omega_{1}-\omega_{2}-\omega_{3})$,

~~$\varDelta=2+3$,
~~$\varDelta\omega=\omega-\omega_{1}=\omega_{2}+\omega_{3}$,
~~$\varDelta{\bf k}={\bf k}-{\bf k}_{1}={\bf k}_{2}+{\bf k}_{3}$,

$d_{1\varDelta}=d1~d{\varDelta}~\delta(\omega-\omega_{1}-\varDelta\omega)\delta
({\bf k}-{\bf k}_{1}-\varDelta{\bf k})$,

$d_{23}=d2~d3 ~\delta(\varDelta\omega-\omega_{2}-\omega_{3})\delta(\varDelta{\bf k}-{\bf k_{2}}-{\bf k_{3}})$,

$E^{+}_1$ is the high frequency wave field,
and
$\delta n/{n_0}$ is the low frequency
variation of plasma density given by
\begin{eqnarray} &&\hspace{0.0cm}\frac{\delta
n}{n_0}=-\frac{1}{4\pi n_{0}T_e}\left(1-\frac{\hbar^2
k^2}{12m^2_ev^2_e}\right)\frac{\varepsilon^{(i)}}{\varepsilon}
\int\frac{{\bf k}_2\cdot{\bf k}_3}{|{\bf k}_2||\bf
{k_3}|}E^{+}_2E^{-}_3 d_{23}.\label{b8}
\end{eqnarray}
\\\\
\noindent{\bf Acknowledgment.} This study was partially supported by the 
Australian Research Council (ARC).

\end{sloppypar}
\end{document}